\documentclass[aps, pra, english,superscriptaddress,floatfix,twocolumn]{revtex4-2}

\usepackage{graphicx}
\usepackage{dcolumn}
\usepackage{bm}
\usepackage{braket}
\usepackage{amsmath, amsthm, amssymb}
\usepackage{algorithmicx}
\usepackage{color}
\usepackage[english]{babel}
\usepackage{caption, subcaption}	
\usepackage{qcircuit}
\usepackage{array}

\newcommand{\mytilde}{\raise.17ex\hbox{$\scriptstyle\mathtt{\sim}$}}

\theoremstyle{definition}

\theoremstyle{definition}

\makeatletter
\DeclareRobustCommand\ket[1]{%
  \@ifnextchar\bra{\k@t{#1}\!}{\k@t{#1}}%
}
\newcommand\k@t[1]{{|{#1}\rangle}}
\makeatother

\begin{document}

\preprint{APS/123-QED}

\title{Near-Term Reduction in Nonlocal Gate Count from Distributed Logical Qubits}

\author{Bruno Avritzer}
    \affiliation{Leidos, Inc.,
        4001 Fairfax Dr, Arlington, VA 22203}
    \email{Bruno.Avritzer@leidos.com}

    \author{Nathan Sankary}
    \affiliation{Leidos, Inc.,
        4001 Fairfax Dr, Arlington, VA 22203}


\date{\today}


\begin{abstract}

Modular quantum computing architectures require error correction schemes that remain effective in the presence of noisy inter-processor operations. As such, minimizing the number of such operations on logical circuits partitioned across quantum processors is a primary objective of distributed quantum computing. In this work, we develop basic techniques for qubit allocation using an exemplar color code family and explore generalizations to other color codes. In particular, we show that a 10\% reduction in processor-nonlocal gates is achievable in a setting where syndrome extraction occurs after every logical gate, as in today's devices, and that this scales to significantly greater advantages in the multi-qubit case. We also explore methods of achieving universal gate sets efficiently in this distributed logical setting and evaluate the trade-offs of multiple approaches such as magic state distillation, code switching, and a new method based on logical swaps. Finally, we discuss some considerations for an allocation algorithm for these architectures to perform scalably and connect it to existing work on quantum circuit partitions.   

\end{abstract}

\maketitle

\section{Introduction}
\label{sec:intro}
Quantum error correction (QEC) is a key ingredient for scalable quantum processing. To date, there have been a wide variety of theoretical approaches to QEC, with myriad trade-offs and features. However, quantum computing hardware has recently advanced to the point of demonstrations of QEC in practice \cite{google-qec, quera-qec}, allowing practical experiments to validate these different approaches. 

This development has brought practical QEC to the forefront, but has also exposed scaling difficulties that are common across a variety of quantum computing realizations. This has motivated developments in distributed quantum computing (DQC)\cite{dqc-ion,modular-trapped-ion,scaling-modular-photonic-qc}, a method of scaling large processors using smaller, interconnected processor modules. This approach, while promising, faces real technical challenges. Trade-offs between interconnect speed and fidelity drive a need for distillation, while entanglement generation rates remain low. Furthermore, DQC requires re-evaluation of quantum processing unit (QPU) architectures, with dedicated communication qubits\cite{usp-dqc} a common architectural consideration. The problem of circuit execution with low-speed or high-loss interconnects\cite{ionq-slow} also requires extensive study. 

Although much work has been done on interconnect technologies and circuit partitions, QEC approaches for distributed systems have been relatively underexplored. The key new degree of freedom arising from the distributed setting is the ability to create error correction codes with logical qubits split across multiple processors\cite{cisco-distributed-color-code,nu-quantum-floquet,bc}. This has two key advantages: a reduction in logical errors\cite{clayton-avritzer,jiang-cosmic-ray-errors} and in required entanglement generation rates, the latter of which is our focus as it alleviates the challenges resulting from poor interconnect quality. 

\begin{figure}
\centering
\includegraphics[width=\columnwidth]{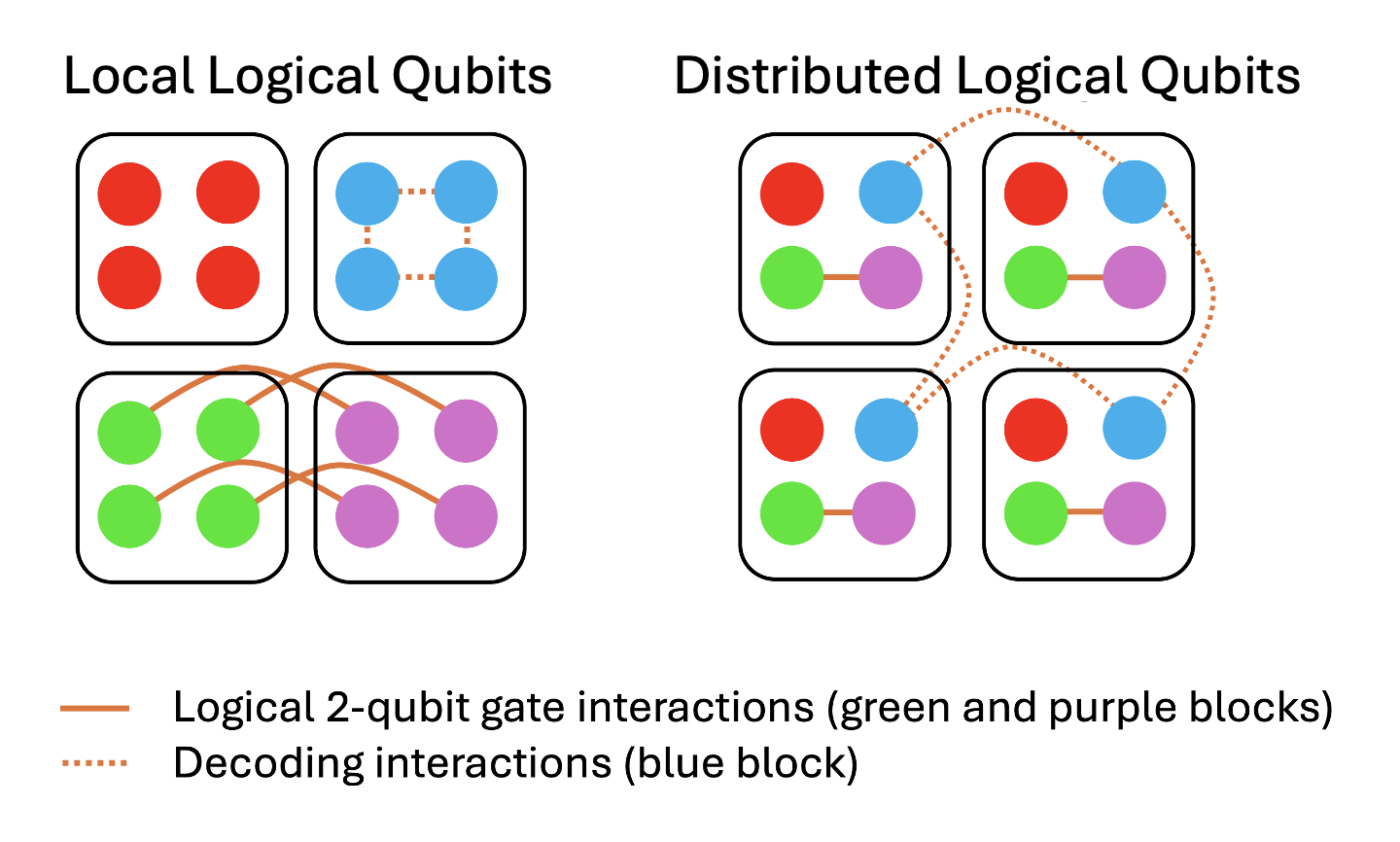}
\includegraphics[width=\columnwidth]{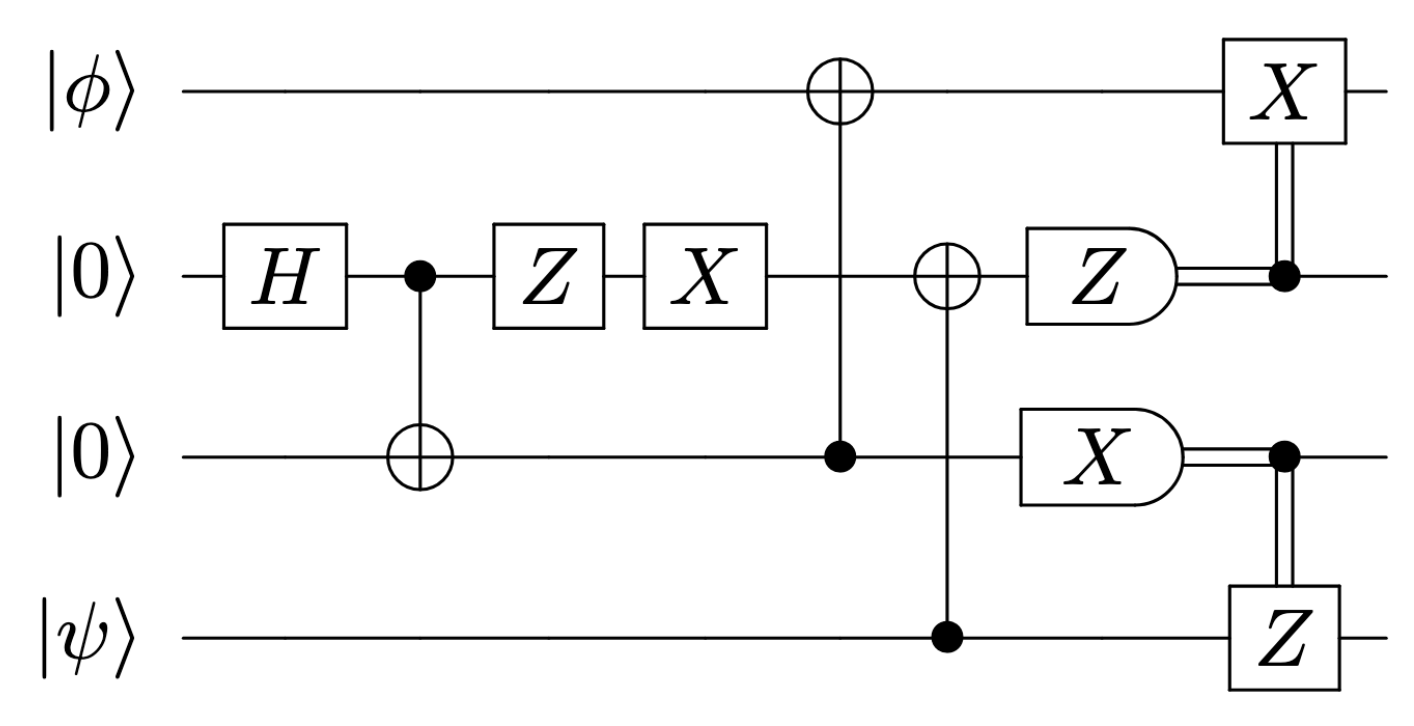}
\caption{Distributed logical qubits require PNL stabilizer measurements as opposed to PNL transversal gate implementations (top). In this work, these PNL gates are mainly implemented by gate teleportation (bottom, $\ket{\psi}$ as the control and $\ket{\phi}$ as the target).}
\label{fig:concept}
\end{figure}

In this work, we consider the trade-off between processor-local (PL) and processor-nonlocal (PNL) operations, with the goal of minimizing PNL operations and their associated entanglement generation overhead. In this work, PNL gates are implemented by CNOT gate teleportation unless otherwise indicated (see Figure \ref{fig:concept}). While it was previously shown in \cite{clayton-avritzer,nkc-arch,vv-dqec} that such gains are possible, in this work we will show that they are achievable even at relatively low qubit counts feasible in today's devices with square-octagon color codes. Furthermore, we will evaluate the performance (in terms of PNL gates) of different approaches to computational universality in this setting, showing that in certain parameter regimes, magic state distillation and code switching can benefit from distribution of logical qubits, both in terms of required qubit overhead and entanglement generation requirements, as well as suggesting a third unconventional approach which may be especially effective in these architectures. To our knowledge, this is the first study of these methods in the distributed logical context. Finally, we will provide an algorithmic template for circuit partitioning based on this structure. These advancements pave the way for concrete realizations of distributed logical qubits, and by extension, realization of distributed logical computations with lower overhead than previously anticipated.

\section{Practical Motivation}
\label{sec:premise}
The setting we consider is distribution of a logical qubit across multiple processors. In this setting, logical operations become processor-local to the extent that the qubits required for a transversal logical CNOT are located on the same module, at the cost of stabilizer measurements becoming processor-nonlocal (again, to the extent that CNOT gates are required between qubits on different modules). The trade-off opens up the question of whether the total number of nonlocal CNOTs can be reduced by such an allocation. The answer, at least for some parameter choices, is yes. If syndrome extractions are infrequent, reduction is possible, but the fidelity of the circuit may be affected. Likewise, there are some scenarios where a small measure of flexibility in the number of qubits allows for allocations which reduce the number of CNOTs. However, these scenarios are not reflective of today's constraints; we will consider only cases where the minimal number of physical qubits (including ancilla) is strictly adhered to, and where one round of syndrome extraction occurs after every logical gate.    

\subsection{Near-Term Advantage on Two Processors}
\begin{figure}
\centering
\includegraphics[width=\columnwidth]{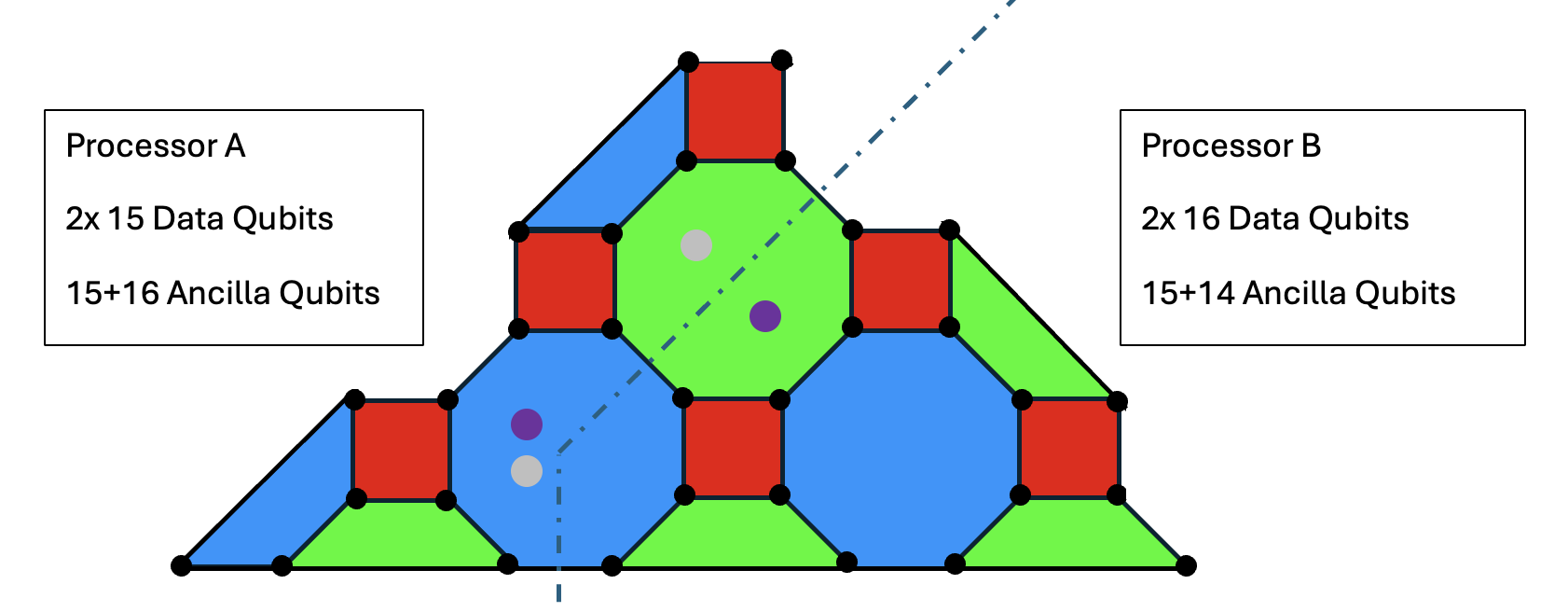}
\caption{The cut which allows an equal partition of the $d=7$ 4.8.8 color code. In this case, there are two 4.8.8 blocks, each with its own set of stabilizer ancillas, and each split 50-50 across two processors. As indicated by the grey and purple dots, in one of the two logical qubits, one ancilla for one of the green octagon checks is allocated to processor B. Since this stabilizer is split 4 to 4, this does not increase the amount of PNL gates, but allows strict adherence to the qubit requirements. The total number of PNL gates is thus 4*7=28, less than the 31 required if the codes are PL.}
\label{fig:example}
\end{figure}

Consider the [[31, 1, 7]] square-octagon color code, which has 15 stabilizer plaquettes. This code requires a total of 61 physical qubits to implement one logical qubit along with its associated syndrome checks. With the split shown in Figure \ref{fig:example}, the number of processor-nonlocal gates required is only 28, contrasted to the usual 31 for ``local'' code allocations, resulting in a reduction of about 10\% per logical gate. One can verify that there is no smaller 2D color code from the 3 standard families (4.8.8, 6.6.6, and 4.6.12) where such an advantage exists. However, it is possible to realize this advantage with the distance-5 square octagon code if some flexibility in processor qubit counts is allowed; if a normal allocation of 33 qubits (17 data+(8+8) syndrome) can be split 40-26, the number of processor-nonlocal gates can be reduced from 17 to 12, or to 16 with a 38-28 allocation. 

Flexibility is very powerful in these scenarios. Although a distance of 7 or 9 is required to see advantage with an exact equality constraint, distance 3 is sufficient without it. For the Steane code, an allocation which puts one qubit from each code on a separate processor allows one to realize two logical qubits with two processors of 24 and 2 qubits, rather than 26, by allocating one physical qubit from each code to the small processor. It is worth noting that in this case the ``reduction in CNOTs'' is somewhat artificial, as there is no alternative allocation that can be made except to allocate ancillas to the small processor, which is clearly suboptimal.

\begin{figure}
    \centering
    \includegraphics[width=\columnwidth]{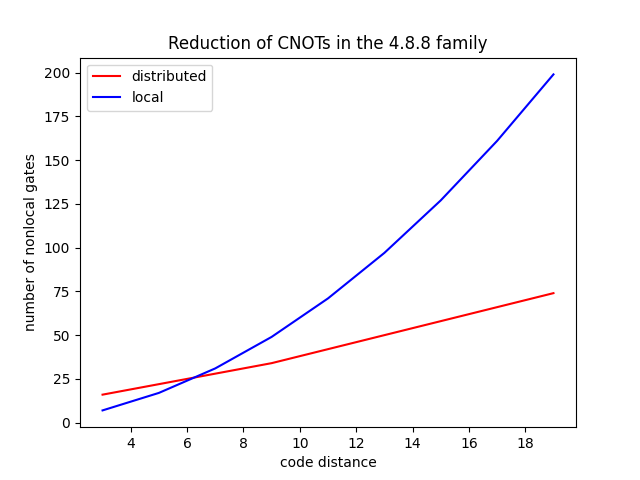}
    \includegraphics[width=\columnwidth]{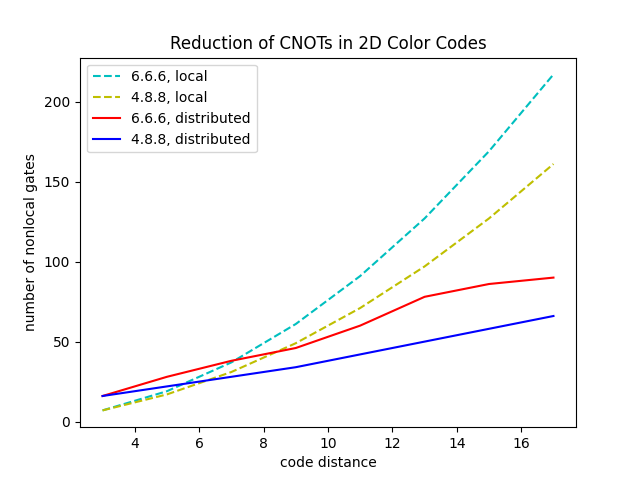}
    \caption{In the two processor (two logical qubit) case (top), in the 4.8.8 family distributed codes see advantage at and above $d=7$, and the overall trend is in agreement with the scaling behavior previously cited. By contrast, the 6.6.6 family of codes sees advantage only at $d=9$ due to the structure of the stabilizers making near-equal bipartitions costly in terms of cut stabilizer weight (bottom).}
    \label{fig:4.8.8trend}
    \end{figure}
Intuitively, the reason for this advantage is clear: the number of PNL gates resulting from stabilizer splits grows with the dividing path across stabilizers and thus linearly with the distance, but the number of nonlocal gates from transversal gate execution is proportional to the total number of qubits in a code block, which grows with the area and thus quadratically in the distance. Numerical results shown in Figure \ref{fig:4.8.8trend} illustrate this trend, with some small deviation from linearity due to the integer constraints, which were formulated as a constraint programming satisfiability (CP-SAT) problem and solved using ORTools\cite{ortools}. It is worth noting that this trend only holds for 2D color codes, but if the code has a $k$-dimensional structure as in higher-dimensional color codes and others, the ratio is still $d^{k}/d^{k-1}=d$. Of course, this comes with a greater base number of PNL gates in the code distance, and as such it is desirable to keep $k$ as low as possible.
\subsection{Alternative Types of Stabilizer Measurement}
There are a few other ways one can imagine to measure stabilizers in the distributed context. One can imagine teleporting, or even swapping (if space and scheduling constraints are too harsh) ancilla qubits across processors. With this approach, one would start with an ancilla on processor A, perform the CNOTs to the relevant qubits that comprise that stabilizer check on that processor, teleport the ancilla to processor B, and then perform the rest of the CNOTs before measuring it. This has the advantage of reducing the cost of high-weight stabilizers to 3 PNL gates required for swaps (only two for teleportation), but risks losing the information in transit. In general this trade-off is not clearly advantageous for color codes with relatively low-weight parity checks, but as we shall later see, this concept provides great advantage when dealing with codes with very high-weight parity checks (such as 3D codes). 
One can also consider fault-tolerant syndrome extraction, namely Shor-, Knill-, and Steane-style methods. Of these, Steane-style seems appealing due to the idea of keeping gates processor-local. However, Steane-style extraction only performs well to the degree that the state preparation circuits can be performed processor-locally. However, depth-optimal preparation of large logical ancilla, especially for given processor topologies, is a challenging question we will not address within the scope of this paper, except to comment that even if, as in \cite{kim}, circuits can be restricted to depth 4, the circuit must be fairly sparse in order to see an advantage. It may be possible to reduce these resource requirements by preparing and verifying the ancilla states using PL constructions, and then distributing the qubits once they are needed for syndrome extraction, but it is not clear that this reduces overhead significantly. Overall, fault-tolerant syndrome measurement poses a challenge for these distributed systems. 
\subsection{More Than Two Processors}
\begin{figure}
    \centering
    \includegraphics[width=\columnwidth]{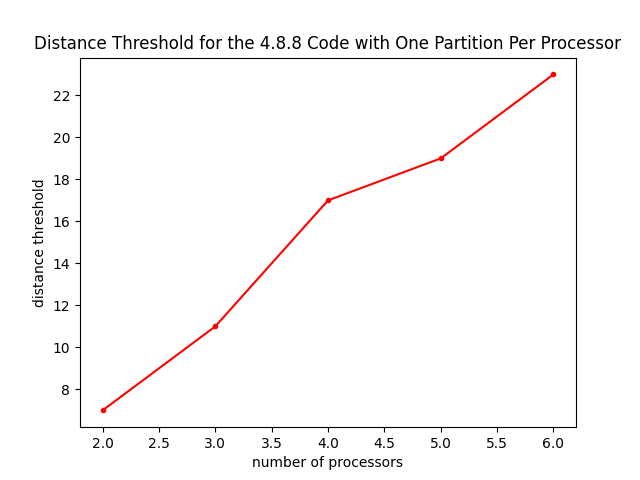}
    \caption{With distribution of a logical qubit block over multiple processors, larger distances (and therefore reductions in PNL CNOTs) may be required to realize advantage if error correction is performed after each logical CNOT. Assuming the logical qubit is cut once per processor resulting in one partition per processor (2 processors require 1 cut resulting in 2 partitions), so that all two-qubit gates are processor-local, the minimum distance threshold required for an overall PNL reduction scales roughly linearly in the number of processors.}
    \label{fig:multipros}
\end{figure}
Since there is an overall scaling of $d$ vs $d^2$, one may think that partitioning a circuit among many processors can reduce the number of PNL gates even more. However, this is misleading in practice. Each additional cut means more stabilizer measurements in each round become PNL, and as a result there are practical limitations on how many cuts may be made advantageously. Circuit structure is also a big factor. Figure \ref{fig:multipros} shows the lowest distance for which the 4.8.8 and 6.6.6 codes can realize advantage across multiple processors, again with syndrome extraction after every logical gate. It is worth noting that none of our methods in this section have considered the encoding and decoding cost of the code, but as there is a reduction per each gate this cost, which is fixed, can be overcome for sufficiently high-depth circuits.

\section{Universality}
\begin{figure}
    \centering
        \begin{subfigure}{\columnwidth}
            \includegraphics[width=.5\textwidth]{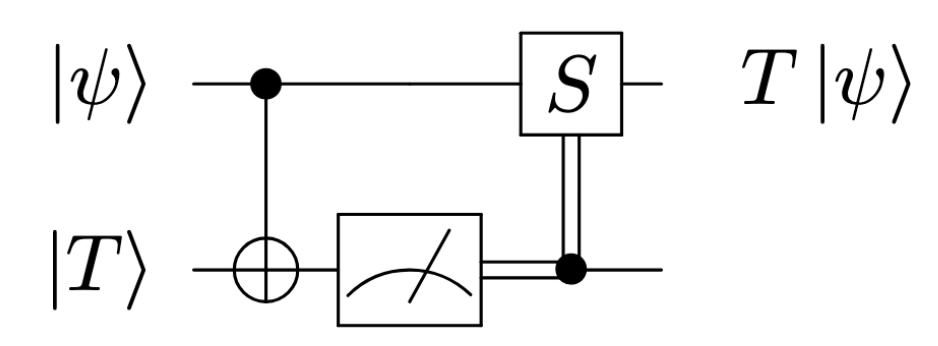}
            \caption{In magic state injection, a high-fidelity magic state $\ket{T}$ is prepared and then consumed to apply a $T$ gate to an input state using the above gadget.}
            \label{fig:msd}
        \end{subfigure}    
        \begin{subfigure}{\columnwidth}
            \includegraphics[width=\textwidth]{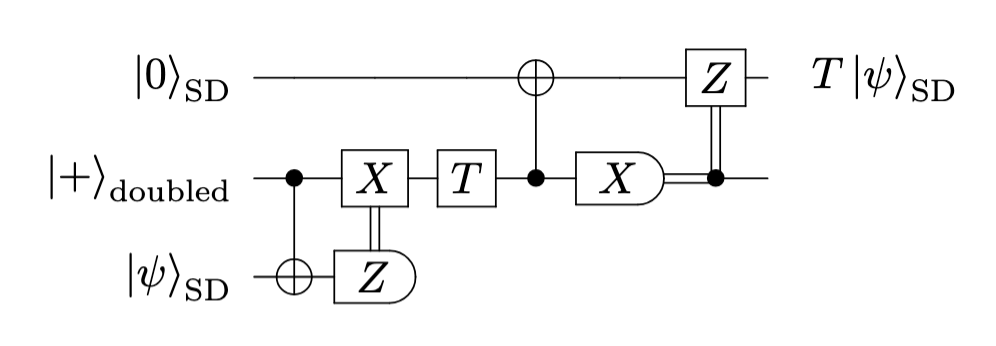}
            \caption{In code switching, a universal gate set is realized transversally by ``switching'' between two codes. The 2D self-dual code realizes $H$ gates and the 3D triorthogonal code realizes $T$ gates. The above gadget realizes a $T$ gate using fault-tolerant code switching between a self-dual code and a doubled, triorthogonal code using transversal CNOTs along only a single face of the 3D code, resulting in extremely low overhead.}
            \label{fig:codeswitch}
        \end{subfigure}
        \caption{Two methods of realizing universality in fault-tolerant systems: magic state injection (a) and code switching (b, adapted from\cite{sullivan}). Although these methods differ in their circuit realizations, in both cases there are regimes where they benefit and suffer from distributed execution across processors.  
    }
    \label{fig:approaches}        
\end{figure}
\label{sec:universality}
We have so far focused on transversally implementable Clifford operations, but in order to do useful computations, significant non-Clifford resources are also required. The Eastin-Knill theorem\cite{eastin-knill}, however, denies the existence of any exact error correcting code with a transversal universal gate set. Nevertheless, there are several methods of performing universal quantum computation with CSS codes. Among the most popular are magic state distillation/injection\cite{bravyi-kitaev-distillation,quera-msd}, where noisy non-Clifford resources are prepared and iteratively refined in quality (``distilled'') before being consumed to apply a non-Clifford operation, and code switching\cite{nokia-switch,steane-reedmuller-switching,heusen}, where a second error correcting code with a transversal non-Clifford gate (often a 3D triorthogonal code) is teleported into, used to apply the non-Clifford gate, and then teleported out of (to the original code). Figure \ref{fig:approaches} illustrates the basic circuit structure of each approach. These two methods have different trade-offs and scaling in the distributed context, but the key trade-off in efficiency is the dimensionality of the stabilizer structure of the underlying codes. Common to both methods, if realized in a distributed fashion, is a reduction in qubit resource requirements: it is no longer necessary to have one magic state factory per processor as each set of distributed code blocks can interact with the magic factory (or code-switching block) on its own. This does, however, require an offset in time of the non-Clifford operations, which may introduce difficulties in circuit compilation for this structure.

\subsection{Magic State Injection}
\label{sec:MSD}
\begin{figure}
    \centering
    \includegraphics[width=\columnwidth]{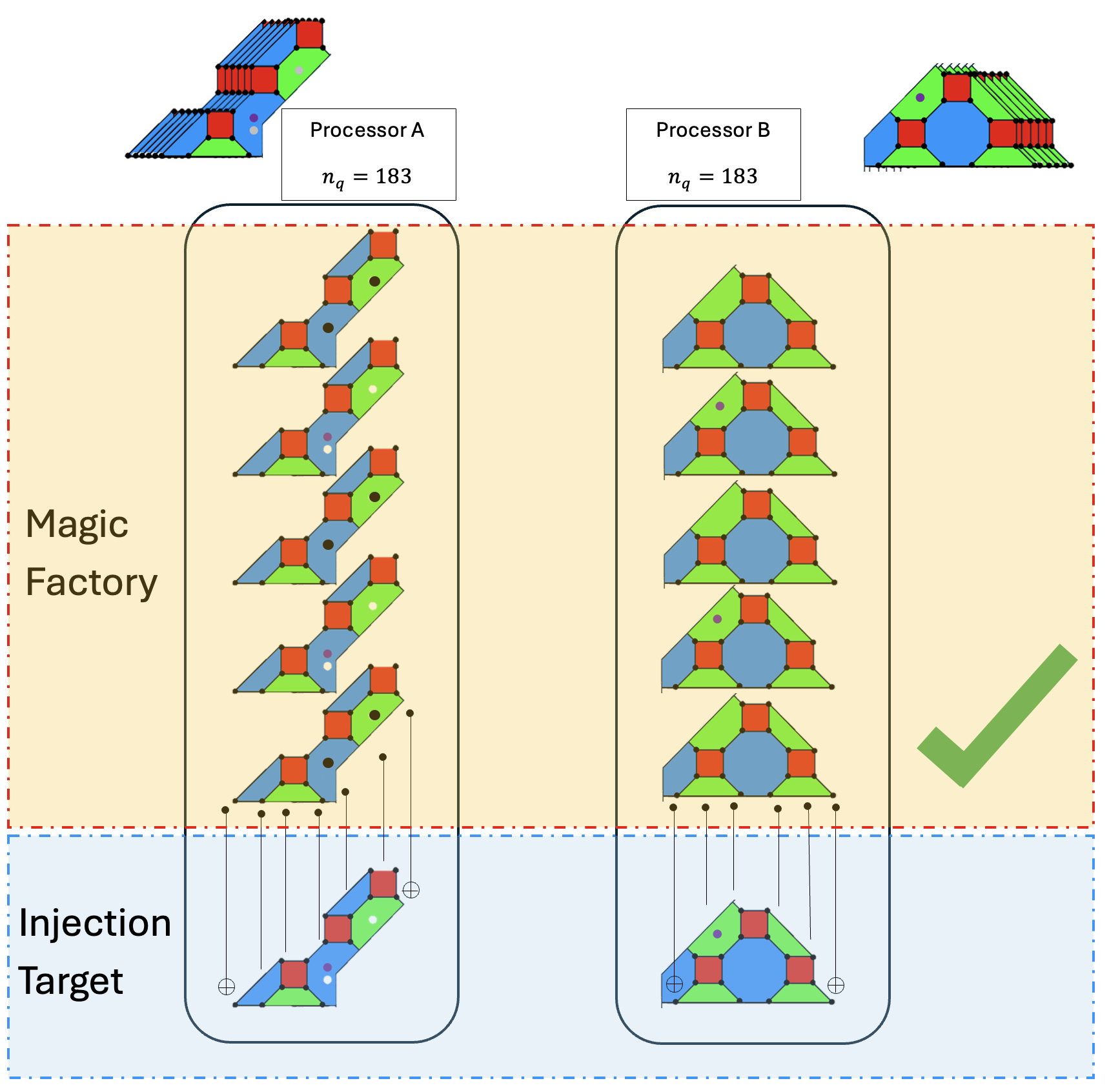}
    \caption{In the fully distributed case, split distillation qubits are allocated on the same processors as split computational qubits. This reduces the per-processor qubit requirements for magic factories while keeping distillation and injection operations PL no matter which qubit is postselected. The total number of PNL gates is $O(d)$, resulting from the syndrome extractions within the individual codes.}
    \label{fig:dmsd}
    \end{figure}
    \begin{figure*}
        \centering
        \begin{subfigure}{.66\columnwidth}
            \includegraphics[width=\textwidth]{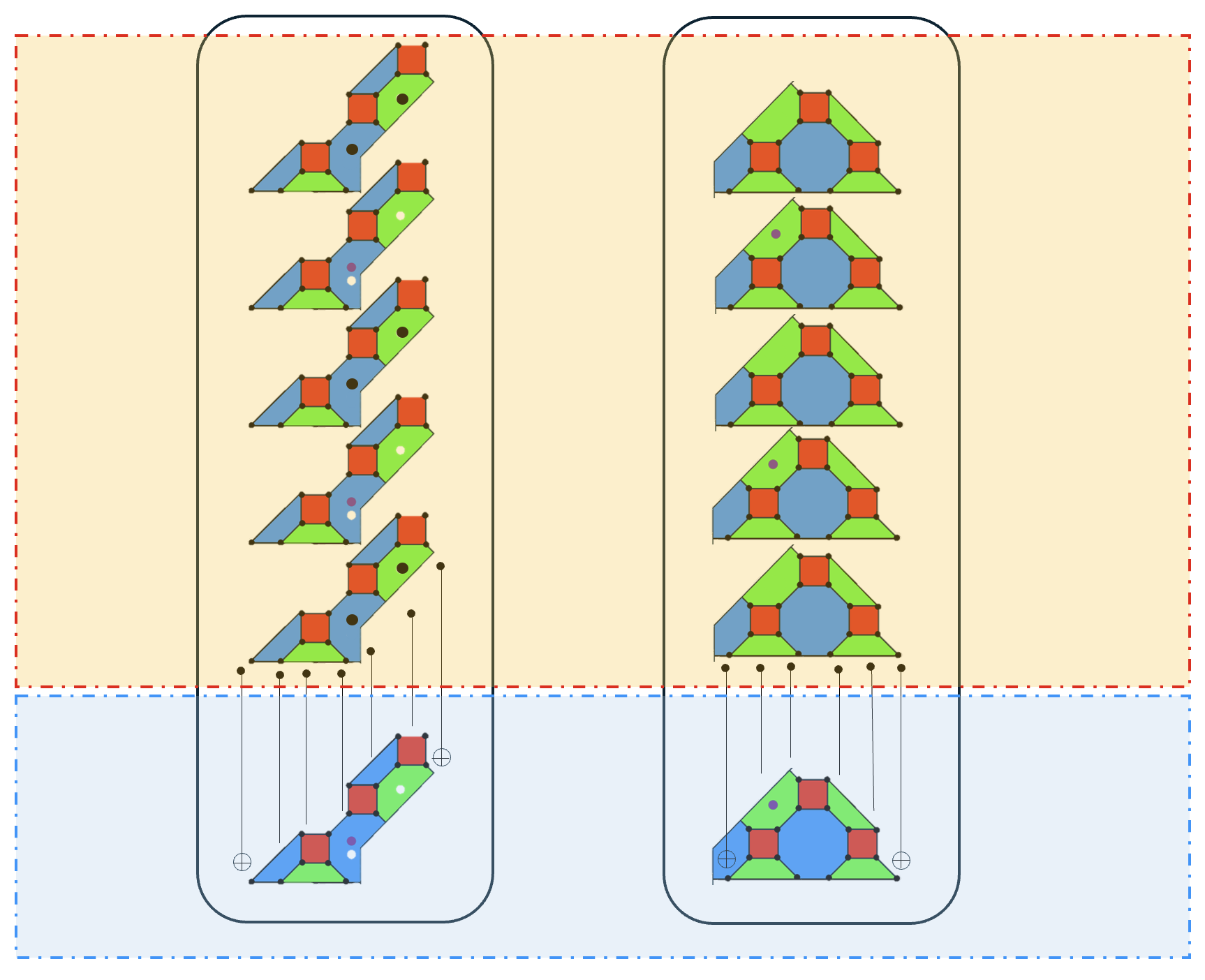}
            \caption{With a fully distributed approach, the total number of logical qubits required is only 3 per processor, and only $O(d)$ PNL gates are required.}
            \label{fig:c1}
        \end{subfigure}
        \begin{subfigure}{.66\columnwidth}
            \includegraphics[width=\textwidth]{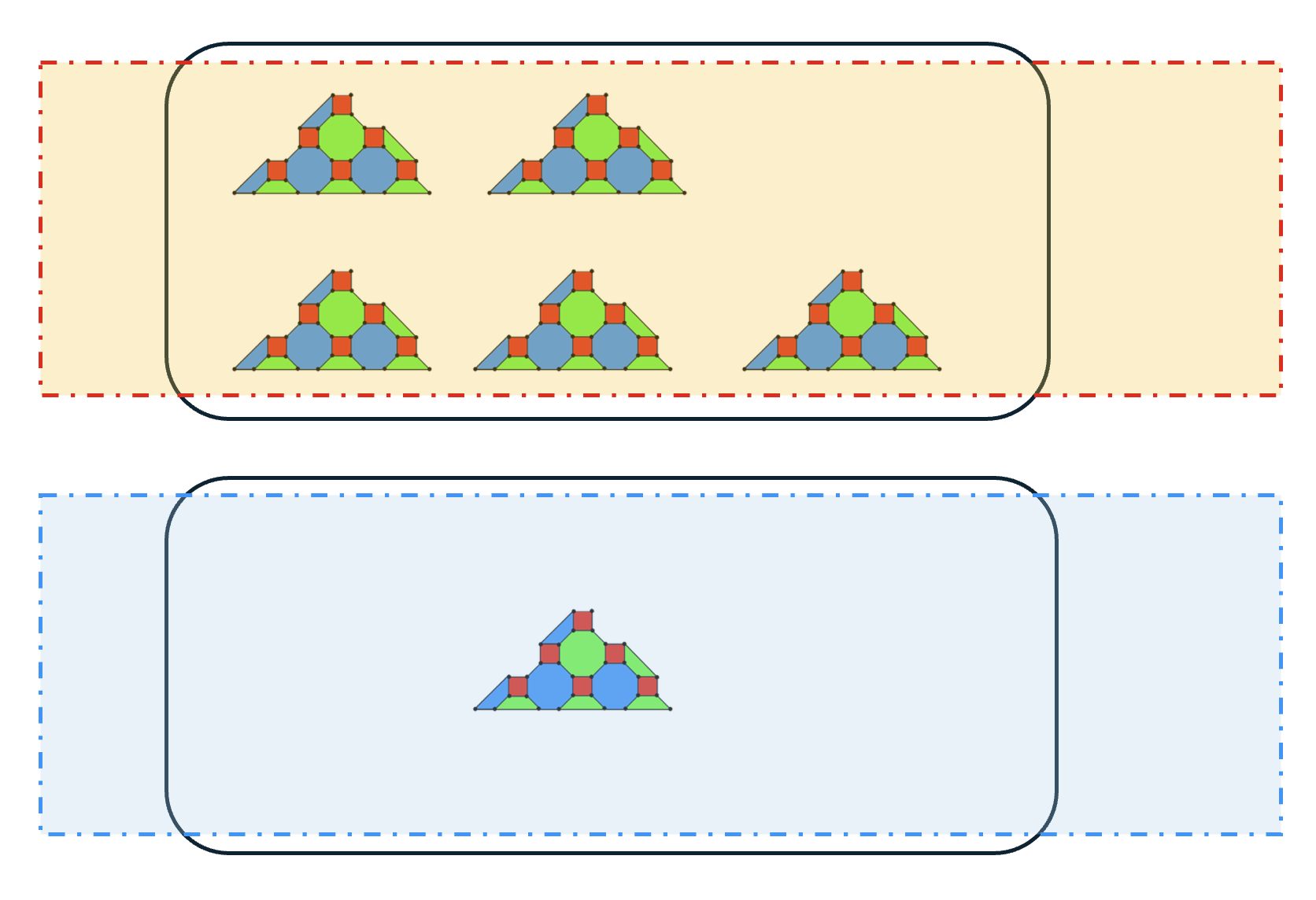}
            \caption{With a fully local approach, 5 logical qubits are required for the distillation processor and $O(d^2)$ PNL gates are required for injection.}
            \label{fig:c2}
        \end{subfigure}
        \begin{subfigure}{.66\columnwidth}
            \includegraphics[width=\textwidth]{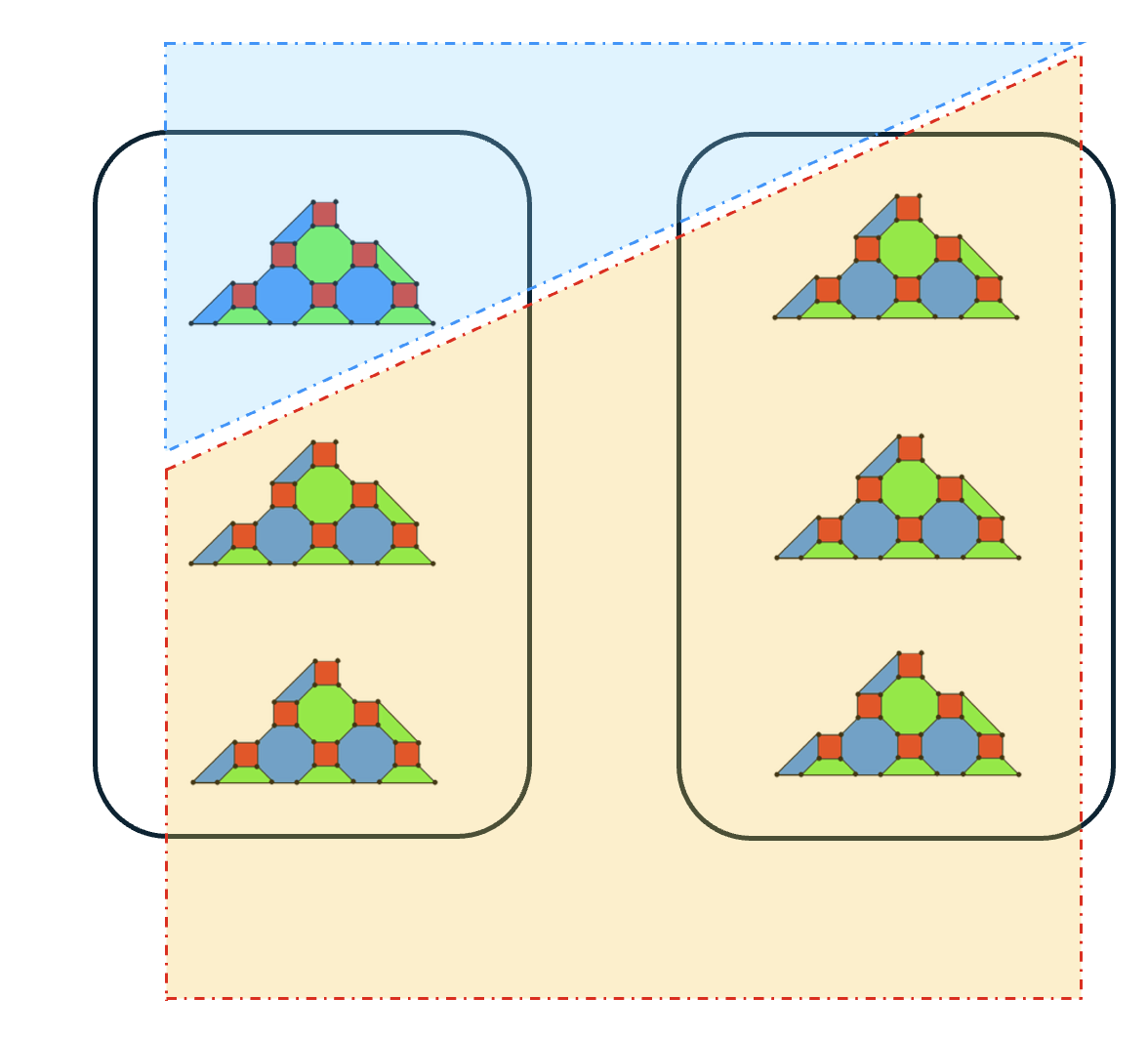}
            \caption{With a mixed approach, 3 qubits are required per processor but the total number of PNL gates is $O(d^2)$.}
            \label{fig:c3}
        \end{subfigure}
        \caption{Comparison of overhead between (a) fully distributed, (b) fully local, and (c) partially local magic factories. The fully distributed factory performs best for small circuits, whereas for large circuits, it has a slightly higher total qubit overhead, but the size of each individual processor can be smaller than that of the magic factory in (b) and a lower number of PNL gates is required compared to (b), although this is network topology-dependent. Approach (c) performs worse in terms of PNL gates, but may have some advantages in meeting network topology constraints.}
        \label{fig:comparison}
    \end{figure*}
5-to-1 logical magic state distillation\cite{bravyi-kitaev-distillation} is comprised of 3 operations: the initialization step which realizes noisy non-Clifford encoded resource states in logical qubits, the refinement step, which consists of decoding circuits for the overarching code (in this case the 5 qubit perfect code\cite{laflamme}) along with postselection and subsequent syndrome extractions within each postselected logical qubit, and the injection circuit which uses this magic state to perform a non-Clifford gate. Of these 3 steps, only the third becomes processor-nonlocal in a distributed setup with an isolated magic factory, but using distributed logical qubits makes only the second step nonlocal. Figure \ref{fig:comparison} shows the relative cost trade-offs of potential realizations. However, this figure does not tell the whole story. Magic state distillation must be executed over many rounds $n_r$ to refine resources at the required infidelity of ~$10^{-8}$ starting from today's physical resources, and although the stabilizer measurement overhead scales only as $d$, the number of such stabilizer measurements increases as $\exp(n_r)$ for a total scaling $O(d\exp(n_r))$. As $d$ is likely on the order of 20 for a fault-tolerant system, this does not allow many rounds of syndrome extraction to be performed in this case to still observe an advantage from the $d$ vs $d^2$ relation, posing a problem for scalability. It does not, however, rule out the advantage of this approach for near-term proof-of-principle experiments, where only 1 or 2 rounds of distillation can be performed, and in fact lowers the network overhead for such demonstrations. 

One may also consider other approaches to magic state injection in the distributed context, such as the 15-to-1 protocol, but while increasing the base overhead may seem a promising method to avoid the scaling resulting from extended rounds of distillation, it is unlikely that this can feasibly be extended to provide advantage in a fault-tolerant regime.

\subsection{Code Switching}
\label{sec:switching}
\begin{figure}
    \centering
    \includegraphics[width=\columnwidth]{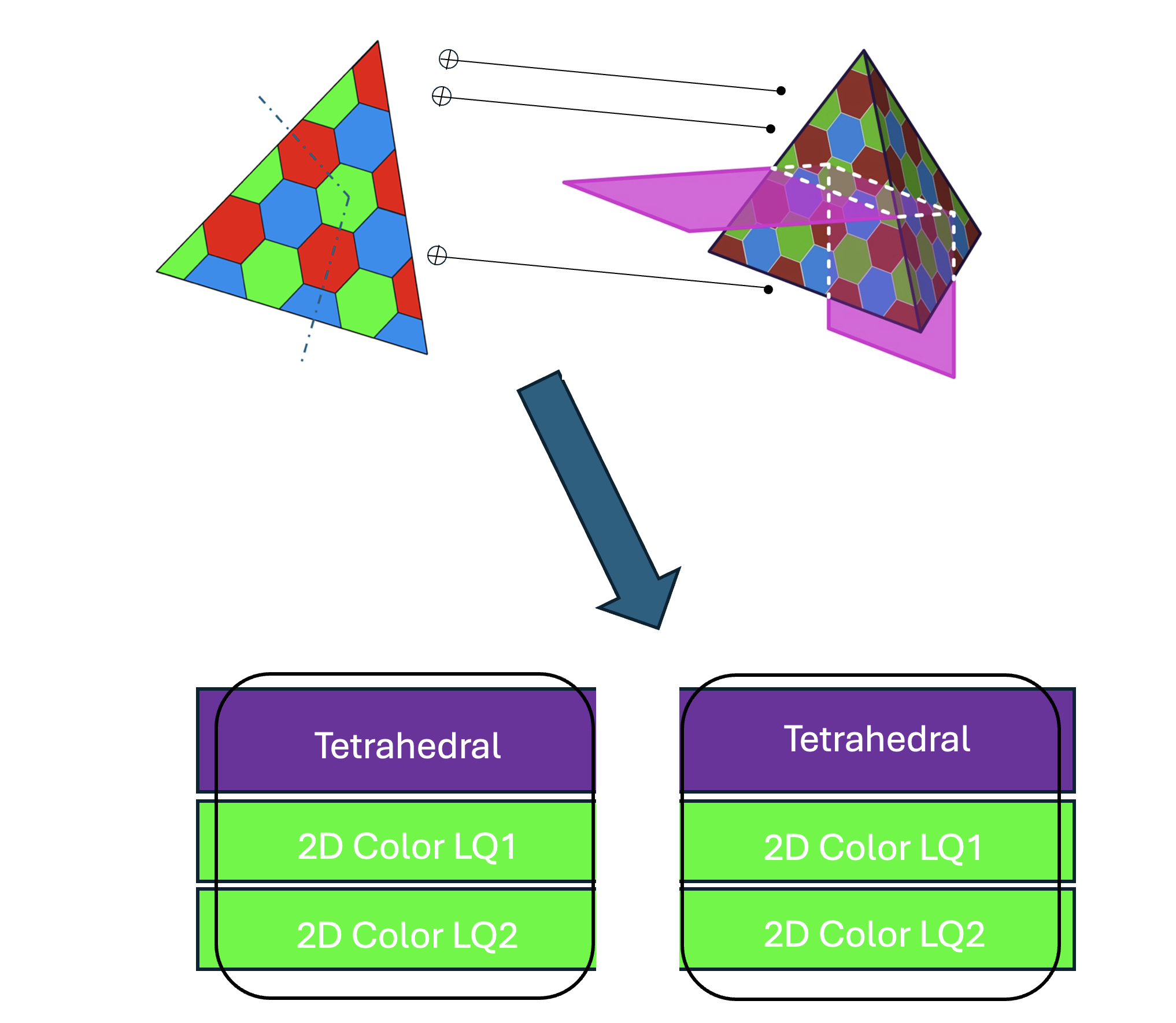}
    \caption{When 3D and 2D codes are both split across processors, code switching may be performed processor-locally. With sufficient connectivity, this can reduce qubit overhead compared to single processors with both codes by allowing multiple logical qubits to interact with the tetrahedral code processor-locally.}
    \label{fig:switching}
\end{figure}

Code switching, especially using triorthogonal codes and self-dual color codes, is another route to universality one can consider. At first, code switching appears to have promising features in the distributed context. The qubit overhead required is significantly smaller than that of MSI, and it retains the benefit of fully processor-local execution with the right architecture. Furthermore, a 3D color code apparently requires overhead $O(d^3)$ for code switching in the local case and only $O(d^2)$ in the distributed case. However, code doubling\cite{code-doubling} provides a complication. The code doubling construction allows for an efficient way of performing non-Clifford gates, with only $O(d^2)$ nonlocal gate cost rather than the expected $O(d^3)$\cite{sullivan}. This cost is reduced to zero if the switched code is colocated with the comptuational qubits, as in Figure \ref{fig:switching}. This construction can be scaled to distance $d$ using a doubled triorthogonal code of distance $d-2$ and a self-dual code of distance $d$ as ingredients, and as such can be implemented recursively to construct codes of arbitrarily high distance with cheap transversal CNOTs. However, since the constructed code always has three-dimensional stabilizer structure, the partition across processors incurs a cut cost of $O(d^2)$ as well in terms of the number of PNL gates required to perform a full round of stabilizer measurements, providing no scaling advantage over local codes. This perspective explains the advantage of MSI as well: the codes involved there were all 2D, and so an advantage was observable at small scales. 

\begin{figure}
    \centering
    \begin{subfigure}{\columnwidth}
        \includegraphics[width=\textwidth]{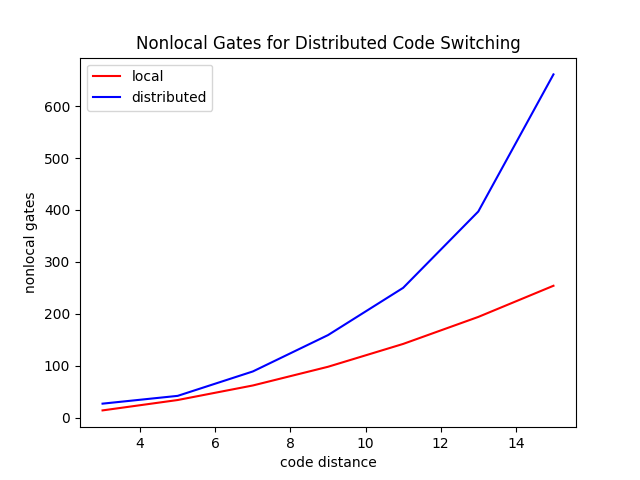}
        \caption{With a distributed approach and standard stabilizer measurements, there is no advantage to distributing the code switching block as opposed to having it local on a separate processor.}
        \label{fig:s1}
    \end{subfigure}
    \begin{subfigure}{\columnwidth}
        \includegraphics[width=\textwidth]{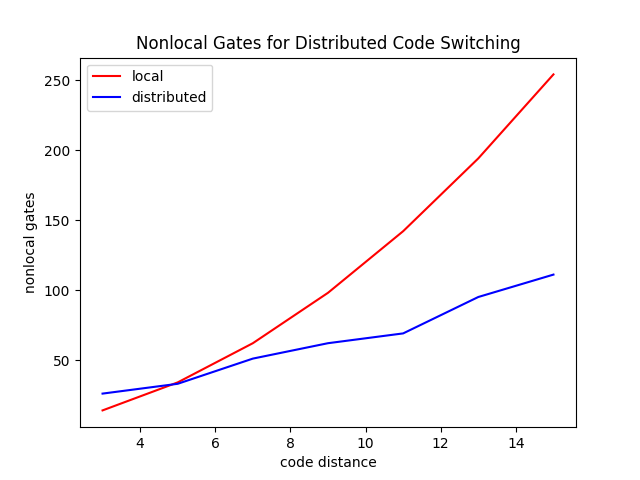}
        \caption{With ancilla swapping/teleportation, it is now possible to observe an advantage from distributing the triorthogonal code switching block.}
        \label{fig:s2}
    \end{subfigure}
    \caption{Comparison of local vs distributed code switching costs for two different forms of stabilizer measurement.}
    \label{fig:switchcost}
\end{figure}

Although there is no scaling advantage in this context as both are $O(d^2)$, this does not rule out a persistent practical advantage. Indeed, we find that there is an advantage under certain scenarios. Figure \ref{fig:switchcost} shows that it is possible to observe an advantage if the previously-discussed ancilla swapping technique is implemented, which reduces the weight of cut stabilizers significantly. This is because the code doubling construction causes cut stabilizers for roughly equal partitions to be both dense and high-weight, and so this technique is necessary to realize an advantage. 

Gauge fixing using, for example, a 3D color code, is a promising approach to implementing non-Clifford gates as well, but ultimately not so different from code switching in resource cost. While it is possible to realize non-Clifford gates using subsystems of the larger code, logical CNOTs across blocks have to be implemented at the higher code level, requiring $O(d^3)$ PNL gates. This dampens the greatest advantage of distributing logical qubits in the first place - a reduction in the number of PNL gates. Another consequence of this is that each logical qubit requires signiicantly more physical qubit overhead. Still, this approach may have some benefit if the amount of non-Clifford gates in the circuit is the most important factor. It may also be possible to reduce the overhead using a construction based on the doubled code, but this is outside the scope of this work. 
\subsection{Dynamic Swaps}
\begin{figure}
    \centering
    \includegraphics[width=\columnwidth]{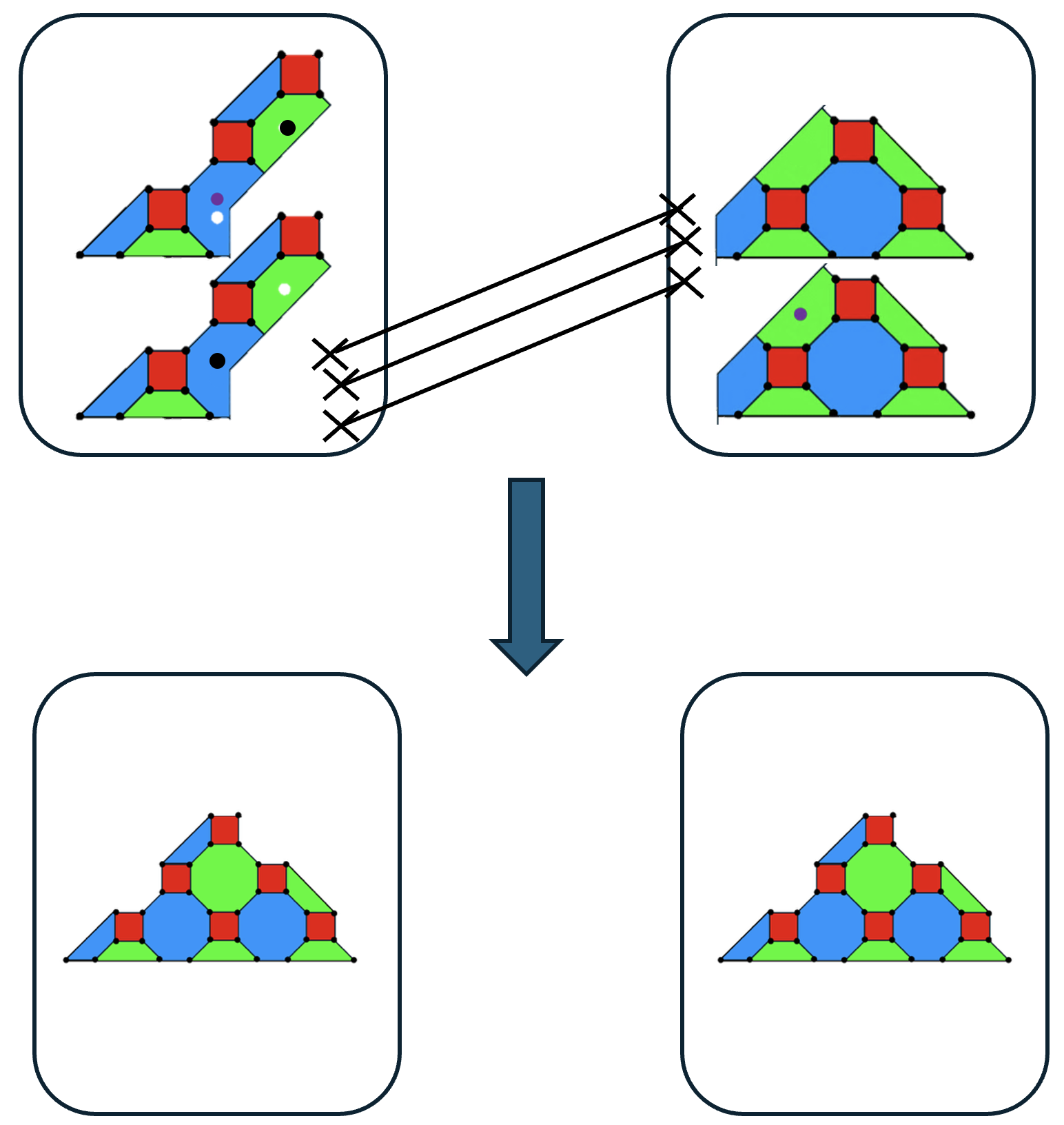}
    \caption{By swapping (data to data) or teleporting (data to ancilla) the qubits from the split code into local structures, non-Clifford resources can be acquired with significantly fewer PNL gates, assuming there is a long chain of switching or distillation required due to circuit gate structure. This can then be undone or partially undone to execute chains of CNOTs, which can then become local as well. }
\label{fig:swapfix}
\end{figure}

There is a final tactic we can use for universality, which bears some similarity to the technique used in \cite{nkc-arch} to implement CNOTs. If there is a long sequence of swaps between H and T gates, for example (which is common in subroutines used for Hamiltonian simulation among other problems) we can simply redefine the code structure as shown in Figure \ref{fig:swapfix}. If processor qubit counts are equal and strictly maintained, it is possible to swap the ancilla and code qubits to different positions using logical swap operations, turning distributed codes into local codes. This allows code switching and magic state distillation to be done processor-locally, but does require magic factories or code switching qubits to be local to the same processor, increasing space overhead. Nevertheless, the reduction in PNL gates resulting from this method can be enormous. Which of these methods proves most fruitful will depend specifically on the processor topology and circuit to be executed.

\section{Circuit Execution}
\label{sec:execution}
\begin{figure*}
    \centering
        \begin{subfigure}{.45\columnwidth}
            \includegraphics[width=\textwidth]{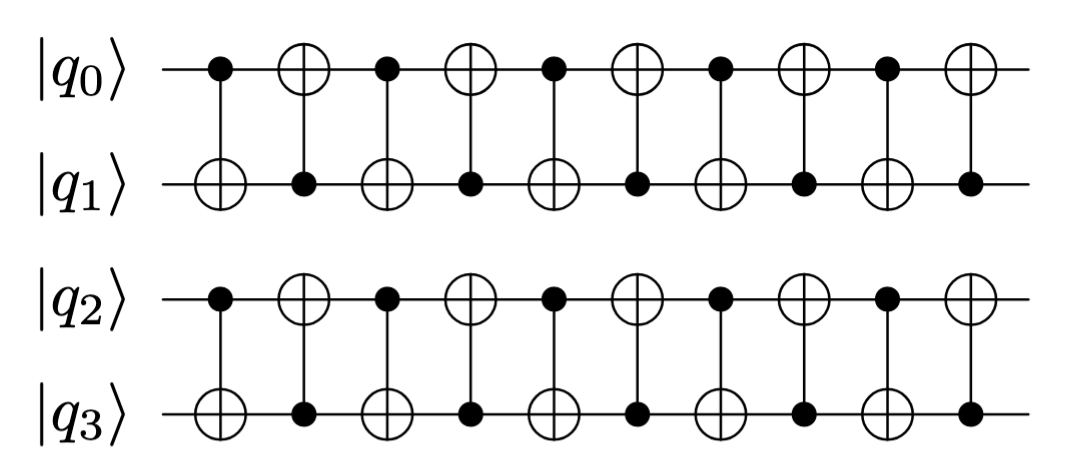}
            \caption{A zero cost circuit which can be partitioned among two two-qubit processors fully locally.}
            \label{fig:counterexample}
        \end{subfigure}    
        \begin{subfigure}{1.5\columnwidth}
            \includegraphics[width=\textwidth]{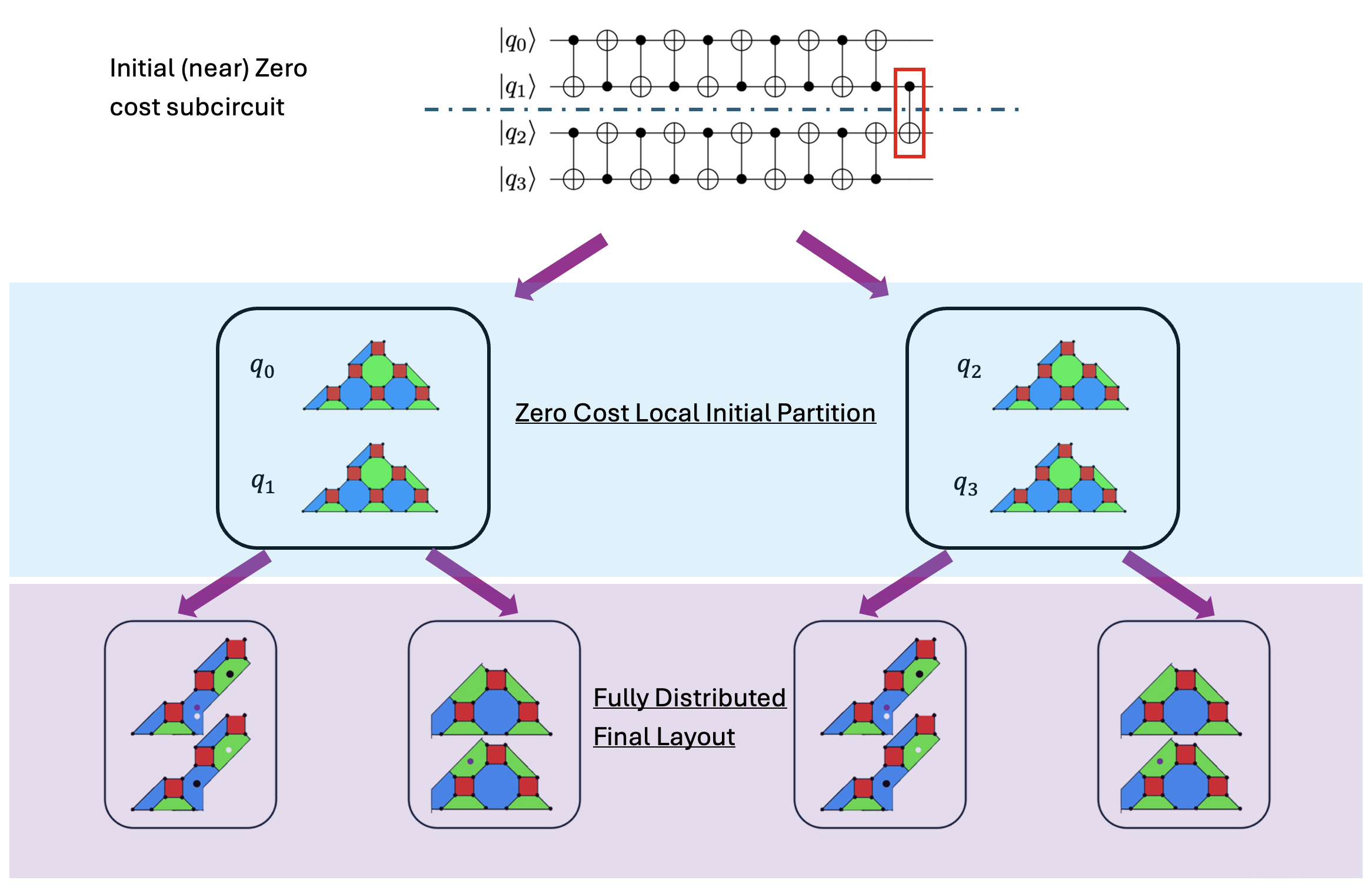}
            \caption{A multilevel partitioning which trades smaller individual processors for more PNL gates, but reduces PNL gates by 10\% compared to 4 processors with one logical qubit each.}
            \label{fig:counterfig2}
        \end{subfigure}
        \caption{The initial partition is optimal for the circuit shown, requiring only 31 PNL gates resulting from the red CNOT. However, if the network cannot accommodate 2 logical qubits per processor, the fully distributed layout allows for an optimized reduction in qubit requirements. In the purple layout, 591 PNL gates are required, compared to 651 with one logical qubit per processor. Because $d$ is small and the circuit structure is able to be partitioned in a way that accommodates this architecture, the $O(d)$ scaling dominates. Note that it is not desirable to split each logical qubit into 4 blocks across all 4 processors, as the gain from making the red gate PL is outweighed by the greater stabilizer measurement cost from more partitions, and that cost must be paid 20 times.}
    \label{fig:counters}        
\end{figure*}

Now that the advantage has been proven and the roadblock of non-transversal gates required for universality has been overcome, one may think that we have an ingredient for fully distributed logical qubit networks. However, this is not necessarily a viable paradigm. Consider a case where one has more than two logical qubits, say 4, and only two large processors (large enough to accommodate 2 code blocks each). We may think to distribute each of these codes once again, but this is likely NOT advantageous. In the case of a logical circuit such as that in Figure \ref{fig:counterexample}, distributing code blocks would make all logical operations processor-local at the cost of nonlocal stabilizer measurement cost, with a seemingly favorable ratio as before. However, it would be possible to simply colocate qubits 0 and 1 on one processor and 2 and 3 on another. This way both logical gates and syndrome extractions would be processor-local. Although the $O(d)$ vs $O(d^2)$ trade-off is correct, it does not tell the whole story; circuit structure and processor constraints are critical factors.

As it turns out, for some circuits and processor configurations, a mixture of fully-distributed and fully-local architectures yields better results. The key determining factor is the circuit statistics. We define 2 relevant parameters that define the ideal architecture and processor size requirements: the ratio of one-qubit transversal gates to two-qubit transversal gates to one-qubit non-transversal gates (requiring MSI or similar approaches) as well as the circuit graph structure for these operations. As an example, if we have at most 3 processors, and the circuit can be broken down so that 99 percent of two-qubit gates are between 3 groups of qubits, the size of these groups constrains the size of the three processors (minus the 1\% which can be allocated at will). If these processors are unrealistically large, they can then be subdivided as the two-qubit gates are processor-local within this large processor, and therefore efficiently implementable using distributed logical qubits, resulting in PNL operations that scale only as $O(d)$. This natural structure makes the problem a good fit for a multilevel partitioning algorithm, such as KaHyPar\cite{kahypar} or KaMinPar\cite{kaminpar}, as the problem breaks down according to zero-cost circuits studied in\cite{zerocost}. As we have seen, adding magic factories in these distributed processors decreases the required number of Bell pairs and the total qubit count but adds a fixed overhead related to the magic state fidelity required. A good rule of thumb is that the percentage of gates which is in the larger partitions (previously 99\%) should be equal to the square root of the number that are not; this way the quadratic and linear costs end up on the same order of magnitude and the total cost is therefore also on the order of magnitude of the fully distributed element of the circuit. It is also possible to run this algorithm iteratively if smaller processors are required or available. The risk is that if there are many non-transversal gates, moving to too small a processor size can significantly increase the total number of PNL gates required from e.g. code switching. As such, the number of rounds of partitioning must be viewed in this light, and the final processor layout is a function of the ratio of the three terms discussed earlier.

\section{Discussion and Future Work} \label{discussion}

In this work, we have placed bounds on the efficiency of distributed compilations resulting from distributed logical color code qubits. These bounds largely result from the cost of nonlocal stabilizer measurements. However, there are several techniques which can alleviate this problem, including Floquetization\cite{nu-quantum-floquet}, flag syndrome extraction\cite{flag}, and more. These techniques could potentially increase the advantage of logical qubit distribution even more than was shown in this paper. 

Furthermore, we have considered only a limited family of codes in this paper -- the color codes -- which leaves open the possibility of other code families which can demonstrate similar advantages. The exploration of other suitable code families is left for future work. 

Finally, allowing for small (or large) deviations in the number of qubits per processor may allow for more scenarios with CNOT reductions. It may also reflect more realistic conditions in a distributed network of heterogeneous processors. We leave these scenarios to be explored in future work. 
\section{Acknowledgements}
BA would like to thank Todd Brun and Colin Bloomfield for helpful discussions. We thank Leidos for funding this research through the Office of Technology and Matt Rudawsky for helping with graphics production. Approved for public
release 26-LEIDOS-0304-30777.
\bibliography{ref}


\end{document}